\documentclass[twocolumn,showpacs,preprintnumbers,amsmath,amssymb,aps,prc]{revtex4-1}

\usepackage{graphicx}
\usepackage{dcolumn}
\usepackage{bm}

\begin{document}

\preprint{APS/123-QED}

\title{Coulomb displacement energies as a probe for nucleon pairing in the $f_{7/2}$ shell}

\author{A.~Kankainen} \email{anu.kankainen@ed.ac.uk}  
\altaffiliation{University of Edinburgh, Edinburgh, EH9 3JZ, United Kingdom}
\affiliation{Department of Physics, University of Jyv\"askyl\"a, P.O. Box 35, FI-40014 University of Jyv\"askyl\"a, Finland}
\author{T.~Eronen}
\altaffiliation{Max-Planck-Institut f\"ur Kernphysik, Saupfercheckweg 1, D-69117 Heidelberg, Germany}
\affiliation{Department of Physics, University of Jyv\"askyl\"a, P.O. Box
 35, FI-40014 University of Jyv\"askyl\"a, Finland}
\author{D.~Gorelov}
\affiliation{Department of Physics, University of Jyv\"askyl\"a, P.O. Box 35, FI-40014 University of Jyv\"askyl\"a, Finland}
\author{J.~Hakala}
\affiliation{Department of Physics, University of Jyv\"askyl\"a, P.O. Box 35, FI-40014 University of Jyv\"askyl\"a, Finland}
\author{A.~Jokinen}
\affiliation{Department of Physics, University of Jyv\"askyl\"a, P.O. Box 35, FI-40014 University of Jyv\"askyl\"a, Finland}
\author{V.S.~Kolhinen}
\affiliation{Department of Physics, University of Jyv\"askyl\"a, P.O. Box 35, FI-40014 University of Jyv\"askyl\"a, Finland}
\author{M.~Reponen}
\affiliation{Department of Physics, University of Jyv\"askyl\"a, P.O. Box 35, FI-40014 University of Jyv\"askyl\"a, Finland}
\author{J.~Rissanen}
\altaffiliation{Nuclear Science Division, Lawrence Berkeley National Laboratory, Berkeley, California 94720, USA}
\affiliation{Department of Physics, University of Jyv\"askyl\"a, P.O. Box 35, FI-40014 University of Jyv\"askyl\"a, Finland}
\author{A.~Saastamoinen}
\altaffiliation{Cyclotron Institute, Texas A\&M University, College Station, TX, 77843-3366, USA}
\affiliation{Department of Physics, University of Jyv\"askyl\"a, P.O. Box 35, FI-40014 University of Jyv\"askyl\"a, Finland}
\author{V.~Sonnenschein}
\affiliation{Department of Physics, University of Jyv\"askyl\"a, P.O. Box 35, FI-40014 University of Jyv\"askyl\"a, Finland}
\author{J.~\"Ayst\"o}
\altaffiliation{Helsinki Institute of Physics, P.O. Box 64, FI-00014 University of Helsinki, Finland}
\affiliation{Department of Physics, University of Jyv\"askyl\"a, P.O. Box 35, FI-40014 University of Jyv\"askyl\"a, Finland}

\date{\today}

\begin{abstract}
Coulomb displacement energies of $T=1/2$ mirror nuclei have been studied via a series of high-precision $Q_\mathrm{EC}$-value measurements with the double Penning trap mass spectrometer JYFLTRAP. Most recently, the $Q_\mathrm{EC}$ values of the $f_{7/2}$-shell mirror nuclei $^{45}$V ($Q_\mathrm{EC}=7123.82(22)$~keV) and $^{49}$Mn ($Q_\mathrm{EC}=7712.42(24)$~keV) have been measured with an unprecedented precision. The data reveal a 16-keV ($1.6\sigma$) offset in the adopted Atomic Mass Evaluation 2012 value of $^{49}$Mn suggesting the need for further measurements to verify the breakdown of the quadratic form of the isobaric multiplet mass equation. Precisely measured $Q_\mathrm{EC}$ values confirm that the pairing effect in the Coulomb energies is quenched when entering the $f_{7/2}$ shell and reaches a minimum in the midshell.
\end{abstract}


\maketitle       
                      

The strong nuclear force is nearly charge independent, thus nuclei close to the $N=Z$ line exhibit exchange symmetry between protons and neutrons. Mirror nuclei have an equal number of nucleons but the proton and neutron numbers are mutually interchanged. They should have similar energy levels assuming that the nuclear force between two protons is identical to the force between two neutrons (charge symmetry). Therefore, the energy difference between the ground states of $\mathrm{T_Z=(N-Z)}/2=\pm1/2$ mirror nuclei ($Q_\mathrm{EC}$ value) should result from the difference in their Coulomb energies once the neutron-proton mass difference ($\Delta_{nH}$) has been taken into acccount \cite{Hei32,Wig37}. Approximating nuclei as uniformly charged spheres, the Coulomb displacement energies ($\mathrm{CDE}=Q_\mathrm{EC}+\Delta_{nH}$) should increase linearly as a function of $\mathrm{Z_{AVG}/A}^{1/3}$, where $\mathrm{Z_{AVG}}$ is the average charge of the mirror pair. The first studies on Coulomb displacement energies between mirror nuclei focused on extracting charge radii from this linear behavior \cite{Bet38,Pea54,Kof58}. It was soon noted that the CDE values exhibit shell and pairing effects \cite{Fee46,Car54,Jan66,Jan66b,Mue75}. Despite various corrections included in theoretical calculations, about 7~\% of the experimental CDE remained unexplained (Nolen-Schiffer anomaly) \cite{Nol69}. The effect of charge-symmetry breaking \cite{Zuk02} or isospin-nonconserving forces \cite{Kan13} have recently been investigated to explain this anomaly. Accurate $Q_\mathrm{EC}$ values of mirror nuclei are also essential for high-precision beta-decay $ft$ values \cite{Sev08,Nav09,Sat12} to be discussed in a separate paper. 

The energy differences between excited states of mirror nuclei in the $f_{7/2}$ shell have been studied intensively via in-beam gamma-ray spectroscopy \cite{Ben07}. The variation of Coulomb energy differences as a function of excitation energy and spin has been interpreted fairly well with large-scale shell-model calculations albeit an anomalously high two-body Coulomb matrix element for the $J=2$ proton pair has to be implemented to match with the observations \cite{Wil03,War06,Ben07}. To date, the $Q_\mathrm{EC}$ values of most of the $f_{7/2}$ shell mirror nuclei have been known with a modest precision of around 10 keV \cite{AME03,AME12}. For comparison, mirror energy differences in $^{45}$V \cite{Ben06} and $^{49}$Mn \cite{OLe97} are known up to 7159 keV ($J^{\pi}=27/2^-$) and 10726 keV ($J^{\pi}=31/2^-$), respectively, with a precision of around 1 keV or better. In this Rapid Communication, we report on the first high-precision $Q_\mathrm{EC}$-value measurements of $^{45}$V and $^{49}$Mn. This work continues the successful programme on mirror nuclei at the Ion Guide Isotope Separator On-Line (IGISOL) facility: $^{23}$Mg \cite{Saa09}, $^{29}$P \cite{Bla14}, $^{31}$S \cite{Kan10S,Bac12}, $^{53}$Co, $^{53}$Co$^m$, $^{55}$Ni, $^{57}$Cu, and $^{59}$Zn \cite{Kan10}. 

\begin{figure}[!tbp]
\includegraphics[width=\columnwidth]{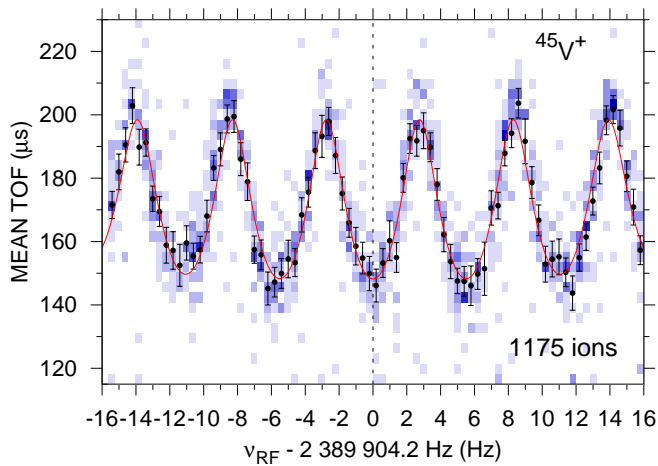}
\caption{A time-of-flight spectrum for $^{45}$V$^+$ using an excitation pattern $25$~ms (On) - $150$~ms (Off) - $25$~ms (On) at JYFLTRAP. Here, the number of ions has been limited to $1-2$ ions/bunch and the TOF window to $115.2-230.4$~$\mu s$.}
\label{fig:TOF}       
\end{figure}

The ions of interest were produced via $^{46}$Ti$(p,2n)$$^{45}$V and $^{50}$Cr$(p,2n)$$^{49}$Mn reactions with a 40-MeV proton beam on enriched $^{46}$Ti and $^{50}$Cr targets at IGISOL \cite{Ays01,Moo13}. They were extracted from the ion-guide gas cell with the help of differential pumping and a sextupole ion guide (SPIG) \cite{Kar08} and subsequently mass-separated with a 55$^\circ$ dipole magnet. The mass-to-charge-separated continuous beam was cooled in the radio-frequency quadrupole cooler (RFQ) \cite{Nie01} and released as short bunches into the JYFLTRAP double Penning trap mass spectrometer \cite{Kol04,Ero12}. The buffer-gas cooling technique \cite{Sav91} was applied in the first trap to select only $^{45}$V$^+$ or $^{49}$Mn$^+$ ions for the high-precision mass measurements using the time-of-flight ion cyclotron resonance technique \cite{Gra80,Kon95} in the second trap. Ion cyclotron excitations were performed via Ramsey's method of time-separated oscillatory fields \cite{Ram90,Bol92,Kre07} with two 25-ms excitation periods separated by a 150-ms waiting time. An example of a time-of-flight spectrum is shown in Fig.~\ref{fig:TOF}. The ion cyclotron resonance frequency is $\nu_c=qB/(2\pi m)$, where $q$ is the charge of the ion, $B$ is the magnetic field in the trap and $m$ the mass of the ion of interest. The magnetic field was calibrated with the corresponding beta-decay daughter nuclei, $^{45}$Ti ($m=44.9581220(9)$~u \cite{AME12}) and $^{49}$Cr ($m=48.9513333(25)$~u \cite{AME12}), which were simultaneously produced with the decay-parent ions. From the measured parent-daughter frequency ratio, the $Q_\mathrm{EC}$ value was calculated as: 
\begin{equation}
\label{eq:qec}
\begin{split}
Q_\mathrm{EC} &= (\frac{\nu_{daughter}}{\nu_{mother}}-1)(m_{daughter}-m_e)c^2~.\\
\end{split}
\end{equation}
With $A/q$ doublets, this will result in a high-precision $Q_\mathrm{EC}$ value even if the daughter nucleus is known only with a modest precision. Additionally, mass-dependent systematic shifts cancel to a level that is well below the statistical uncertainty quoted in this work \cite{Gab09}.

The cyclotron resonance frequencies were fitted with the theoretical lineshape \cite{Kon95,Geo07,Kre07} (see Fig.~\ref{fig:TOF}). The measured frequencies were corrected for the count-rate-effect \cite{Kel03} whenever possible, similarly as in Ref.~\cite{Kank13}. In order to take into account fluctuations in the magnetic field, a correction of $\delta_B(\nu_{ref})/\nu_{ref}=5.7(8)\times 
10^{-11}\text{min}^{-1}\Delta t$, where $\Delta t$ is the time between the two reference measurements, was quadratically added to the statistical uncertainty of each frequency ratio. The cyclotron frequency measurements were carried out interleavedly to reduce the time between the reference measurements. After one frequency scan on the reference ion, 2-5 scans were done for the ion of interest, and the pattern was repeated. Altogether the data were split to 18 frequency pairs of $\nu(^{45}$Ti$^+)/\nu(^{45}$V$^+$) and 20 pairs of $\nu(^{49}$Cr$^+)/\nu(^{49}$Mn$^+$) to obtain resonance quality as in Fig.~\ref{fig:TOF}. The weighted mean of the measured frequency ratios was calculated and used as the final value. The ratio of the outer and inner errors \cite{Bir32} was less than 1 for the data sets, and the larger, inner error was taken as the error of the mean. 

\begin{figure}[!tbp]
\includegraphics[width=\columnwidth]{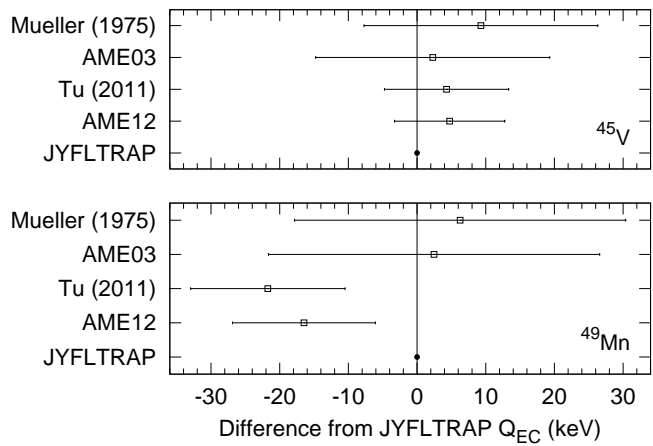}
\caption{$Q_\mathrm{EC}$ values of $^{45}$V and $^{49}$Mn from previous measurements compared to the value determined at JYFLTRAP. Values from the $^{50}$Cr$(p,^{6}$He$)^{45}$V (Mueller \emph{et al.} \cite{Mue75}) and CSRe storage ring (Tu \emph{et al.} \cite{Tu11a}) experiments as well as from the Atomic Mass Evaluations 2003 (AME03) \cite{AME03} and 2012 (AME12) \cite{AME12} all agree well with JYFLTRAP for $^{45}$V. For $^{49}$Mn, the CSRe value \cite{Tu11a}, and consequently the AME12 value, deviate from JYFLTRAP significantly.}
\label{fig:comparisons}
\end{figure}

Table~\ref{tab:results} shows the frequency ratios, $Q_\mathrm{EC}$ and mass-excess values determined in this work. Compared to the previous experiments and mass tables (see Fig.~\ref{fig:comparisons}), JYFLTRAP provides an improvement in precision of about 20~times for the $Q_\mathrm{EC}$ values. Previously, the masses of $^{45}$V and $^{49}$Mn have been determined via $^{50}$Cr$(p,^6$He$)^{45}$V and $^{54}$Fe$(p,^6$He$)^{49}$Mn reaction $Q$ values at the Michigan State University \cite{Mue75}. Recently, they have also been measured with the isochronous mass spectrometry method at the experimental cooler storage ring CSRe in Lanzhou \cite{Tu11a}. The former measurements were the main contributors to the Atomic Mass Evaluation 2003 (AME03) whereas the latter experiment has influenced the new Atomic Mass Evaluation 2012 (AME12) values significantly. As can be seen from Fig.~\ref{fig:comparisons}, all experiments agree with each other in the case of $^{45}$V. For $^{49}$Mn, the $(p,^6$He$)$ and the AME03 values agree well with JYFLTRAP. On the contrary, the CSRe result is $1.9\sigma$ lower than the JYFLTRAP value. This is reflected in the AME12 value, which deviates by $1.6\sigma$ from the JYFLTRAP value. Similar deviations are observed with the mass-excess value of $^{49}$Mn. If the more exotic isobar $^{49}$Fe measured also at CSRe \cite{Zha12} has a similar $1.9\sigma$ shift, then the cubic coefficient of the Isobaric Multiplet Mass Equation (IMME) would agree with zero at $A=49$. Thus, further measurements are essential to confirm the breakdown of the IMME in the $fp$ shell \cite{Zha12}.

\begin{table*}
\caption{\label{tab:results} Frequency ratios $r$, $Q_\mathrm{EC}$ values and mass-excess values $\Delta$ compared to the AME12 values \cite{AME12}.}
\begin{ruledtabular}
\begin{tabular}{ccccccc}
Nuclide & Reference & $r$ & $Q_\mathrm{EC}$ (keV) & $Q_{EC,AME12}$ (keV) &$\Delta$ (keV) & $\Delta_{AME12}$ (keV)\\ 
\hline
$^{45}$V & $^{45}$Ti & $1.0001701102(51)$ & $7123.82(22)$ & $7129(8)$ & $-31885.3(9)$ & $-31881(8)$\\
$^{49}$Mn & $^{49}$Cr & $1.0001691419(52)$ & $7712.42(24)$ & $7696(10)$ & $-37620.3(24)$ & $-37637(10)$\\
\end{tabular}
\end{ruledtabular}
\end{table*}

\begin{figure}[!tbp]
\includegraphics[width=\columnwidth]{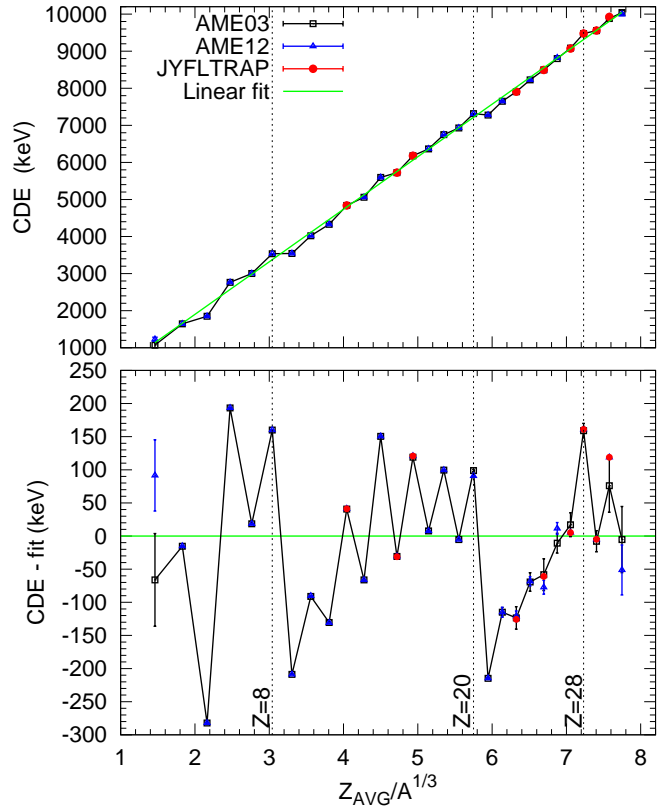}
\caption{Linear fit to the CDE values of the $T=1/2$ mirror nuclei based on the AME03 \cite{AME03} (upper panel) and differences from the fit (lower panel). JYFLTRAP measurements are shown as red dots.}
\label{fig:CDE}       
\end{figure}

To study the impact of JYFLTRAP measurements on the CDE values, an error-weighted linear fit on the CDE values based on the AME03 \cite{AME03} was performed (see Fig.~\ref{fig:CDE}). The shell and pairing effects in the Coulomb energies of mirror nuclei are clearly seen in the lower panel of Fig.~\ref{fig:CDE} showing differences from the linear fit. As the nuclear force is much stronger than Coulomb repulsion, protons tend to pair to angular momentum $J=0$. Thus, mirror nuclei with an even number of protons have spatially closer proton orbitals and therefore higher than average Coulomb energies. At magic proton numbers, this effect is magnified. The magnitude of the odd-even staggering (pairing effect) is substantially reduced when entering the $f_{7/2}$ shell. Recently, Kaneko \emph{et al.} \cite{Kan13} have explained the decrease using isospin-nonconserving isovector and isotensor nuclear interactions in their large-scale shell-model calculations and have compared the results to the AME03 values \cite{AME03} updated with the storage-ring data from CSRe \cite{Tu11a}. The JYFLTRAP data show a fairly similar behavior to AME03 with most striking difference being the stronger staggering above $Z=28$. The AME12 value for $^{49}$Mn is 16-19 keV lower than the AME03 and JYFLTRAP values and would therefore imply a stronger odd-even staggering in the $f_{7/2}$ shell. 

\begin{figure}[!tbp]
\includegraphics[width=\columnwidth]{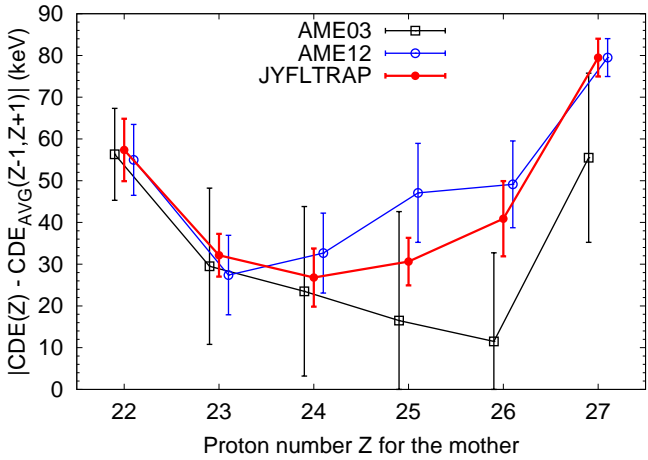}
\caption{Pairing effect in the $f_{7/2}$ shell based on the AME03 \cite{AME03}, AME12 \cite{AME12} and JYFLTRAP values for $^{45}$V, $^{49}$Mn, $^{53}$Co, and $^{55}$Ni (using AME12 values for the other nuclei).}
\label{fig:pairing}
\end{figure}

To illustrate the odd-even staggering of the CDE values in the $f_{7/2}$ shell, its magnitude $|\mathrm{CDE(Z)-[CDE(Z-1)+CDE(Z+1)]/2}|$ has been plotted in Fig.~\ref{fig:pairing}. JYFLTRAP data for $^{45}$V, $^{49}$Mn, $^{53}$Co and $^{55}$Ni (using the AME12 values for the remaining data points) suggest a different trend in the pairing effect along the $f_{7/2}$ shell than the AME03 and AME12: the minimum is reached at the midshell where mirror nuclei show more collective features and deformation. In the midshell, the ground-state spins are no longer $7/2^-$ but correspond to deformed Nilsson orbitals $K=3/2$, $3/2^-$ ($^{47}$Cr) and $K=5/2$, $5/2^-$ ($^{49}$Mn and $^{51}$Fe), see e.g. shell-model calculations \cite{Mar97,Pov99,Bra09}. High quadrupole deformation parameters for $^{45}$V ($\beta_2=0.233$ \cite{Sat12}), $^{47}$Cr ($\beta_2=0.276$ \cite{Sat12}), $^{49}$Mn ($\beta_2=0.216$ \cite{Mol95}), and $^{51}$Fe ($\beta_2=0.198$ \cite{Mol95}) have also been calculated e.g. with the FRDM \cite{Mol95} and EDF \cite{Sat12} models. The experimental CDE values demonstrate nicely that pairing energies are reduced in midshells for deformed nuclei and show peaks at the closed shells \cite{Jan03}.

\begin{figure}[!tbp]
\includegraphics[width=\columnwidth]{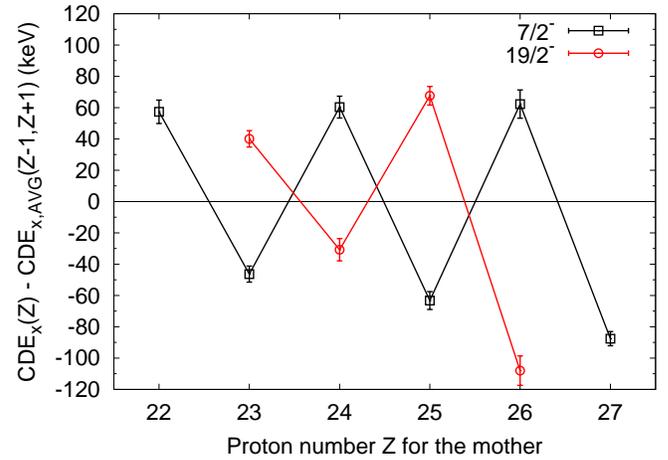}
\caption{Odd-even staggering of CDE values for $7/2^-$ ($J=0$) and $19/2^-$ ($J=6$) states in the $f_{7/2}$ shell.}
\label{fig:Jpairing}
\end{figure}

The change in deformation and ground-state configuration have an influence on pairing. Thus, the pairing effect in the CDE values was also studied for the lowest $7/2^-$ and $19/2^-$ states by combining the CDE data with the mirror energy differences (MED) between the excited states: $\mathrm{CDE_x=CDE+MED_x}$ \cite{NDS43,NDS45,NDS47,NDS49,NDS51,NDS53,Kan10} (see Fig.~\ref{fig:Jpairing}). As a result, the pairing effect is around 60 keV for the $7/2^-$ states throughout the shell.  When the protons are coupled to $J=0$ ($7/2^-$ states), the even-$Z$ nuclei have higher CDEs than their odd-$Z$ neighbors, but an opposite behavior is observed for the maximal angular momentum coupling $J=6$ ($19/2^-$ states). Two protons coupled to $J=0$ are likely to be close to each other, whereas for $J=6$ they tend to be further apart as illustrated in Fig.~2 of Ref.~\cite{War06}. Even these subtle changes in the charge distribution are reflected in the CDE values.

In conclusion, the $Q_\mathrm{EC}$ values of $^{45}$V and $^{49}$Mn have been determined with a superior precision at the JYFLTRAP double Penning trap. The observed $1.9\sigma$ difference to the CSRe measurement of $^{49}$Mn suggests that the masses of the $T=3/2$ isobaric quartets should be remeasured to verify the breakdown of the quadratic form of the IMME in the $f_{7/2}$ shell \cite{Zha12}. Precise Penning trap measurements have made it possible to explore details in the experimental Coulomb displacement energies at a totally new level. The pairing effect in Coulomb displacement energies corresponding to the ground states in the $f_{7/2}$ shell has been shown to be quenched and reaches the minimum at the midshell. Further $Q_\mathrm{EC}$ measurements on $^{43}$Ti, $^{47}$Cr and $^{51}$Fe are anticipated to reach sub-keV precision in the CDE values for all $f_{7/2}$-shell mirror nuclei.

\begin{acknowledgments}
This work has been supported by the EU 6th Framework programme ``Integrating Infrastructure Initiative - Transnational Access", Contract Number: 506065 (EURONS) and by the Academy of Finland under the Finnish Centre of Excellence Programme 2006-2011 (Nuclear and Accelerator Based Physics Programme at JYFL). A.K. acknowledges the support from the Academy of Finland under the project 127301. 
\end{acknowledgments}

\end{document}